\documentclass{birkart}

\usepackage{amssymb}

 \newtheorem{thm}{Theorem}[section]
 
 \newtheorem{lem}[thm]{Lemma}
 
 \theoremstyle{definition}
 
 \theoremstyle{remark}
 \newtheorem{rem}[thm]{Remark}
 
 \numberwithin{equation}{section}
\newcommand{\C}{\mathbb{C}}
\newcommand{\N}{\mathbb{N}}
\newcommand{\R}{\mathbb{R}}
\newcommand{\D}{\mathrm{d}}
\newcommand{\e}{\mathrm{e}}

\newcommand{\HH}{\mathcal{H}}
\newcommand{\QQ}{\mathcal{Q}}
\newcommand{\UU}{\mathcal{U}}
\newcommand{\Hab}{H_{\alpha, \beta}}
\newcommand{\eab}{\epsilon _{\alpha \beta , k}}

\begin{document}
%
\title[Relations between stable and Zeno dynamics]
 {On relations between stable and Zeno dynamics in a leaky graph
 decay model}
\author[P.~Exner]{Pavel Exner}

\address{%
Department of Theoretical Physics, Nuclear Physics Institute,
Academy of Sciences, 25068 \v{R}e\v{z} near Prague, Czechia, and
Doppler Institute, Czech Technical University, B\v{r}ehov{\'a} 7,
11519 Prague, Czechia}

\email{exner@ujf.cas.cz}

\author[T.~Ichinose]{Takashi Ichinose}
\address{Department of Mathematics, Faculty of Science, Kanazawa
University, Kanazawa 920-1192, Japan}
\email{ichinose@kappa.s.kanazawa-u.ac.jp}

\author[S.~Kondej]{Sylwia Kondej}
\address{Institute of Physics, University of Zielona G\'{o}ra, ul.
Szafrana 4a, 65246 Zielona G\'{o}ra, Poland}
\email{skondej@proton.if.uz.zgora.pl}


\date{February 16, 2005}

\begin{abstract}
We use a caricature model of a system consisting of a quantum wire
and a finite number of quantum dots, to discuss relation between
the Zeno dynamics and the stable one which governs time evolution
of the dot states in the absence of the wire. We analyze the weak
coupling case and argue that the two time evolutions can differ
significantly only at times comparable with the lifetime of the
unstable system undisturbed by perpetual measurement.
\end{abstract}

\maketitle

\section{Introduction}

It is well known that the decay of an unstable system can be
slowed down, or even fully stopped in the ideal case, if one
checks frequently whether the system is still undecayed. The first
proper statement of this fact is due to Beskow and Nilsson
\cite{BN} and a rigorous mathematical proof was given by Friedman
\cite{Fr}, but it became popular only after Misra and Sudarshan
\cite{MS} invented the name ``quantum Zeno effect'' for it. In
recent years this subject attracted a new wave of interest -- a
rich bibliography can be found, e.g., in \cite{FM, Sch}.

The motivation of this interest is twofold. On one hand the
progress in experimental methods makes real the possibility to
observe the effect as a phenomenon really existing in the nature,
and ultimately to make use of it. On the other hand, the problem
presents also interesting mathematical challenges. The most
important among them is obviously the question about the
\emph{quantum Zeno dynamics:} if the perpetual measurement keeps
the state of the system within the Hilbert space associated with
the unstable system, what is then the time evolution of such a
state? Some recent results \cite{EI, EINZ} give partial answers to
this question, which we shall describe below, and there are
counterexamples \cite{MaS}, see also \cite[Rem.~2.4.9]{Ex}, which
point out the borders beyond which it has no sense.

In this note we are going to address a different question. Suppose
that at the beginning the interaction responsible for the decay is
absent, so state vectors evolve within the mentioned space which
we below call $P\HH$. Switching the interaction with the
``environment'' in, we allow the system to decay which means
the state vectors may partially or fully leave the space $P\HH$.
If we now perform the Zeno-style monitoring, the system is forced
to stay within $P\HH$ and to evolve there, but what is in this
case the relation of its dynamics to the original ``stable'' one?

A general answer to this question is by no means easy and we do
not strive for this ambitious goal here. Our aim is to analyze a
simple example which involves a Schr\"odinger operator in
$L^2(\R^2)$ with a singular interaction supported by a line and a
finite family of points \cite{EK}. This model is explicitly
solvable and can be regarded as a caricature description of a
system consisting of a quantum wire and dots which are not
connected mutually but can interact by means tunneling. The main
result of this paper given in Theorem~\ref{compar} below is that
in the model \emph{the two dynamics do not differ significantly
during time periods short at the scale given by the lifetime of
the system unperturbed by the perpetual observation.}

Let us briefly summarize the contents of the paper. First we
recall basic notions concerning Zeno dynamics; we will prove the
needed existence result in case when the state spece of the
unstable system has a finite dimension. Sections 3--5 are devoted
to the mentioned solvable model. We will introduce its Hamiltonian
and find its resolvent. Then we will show that in the
``weak-coupling''  case when the points are sufficiently far from
the line the model exhibit resonances, and in Sec.~5 we will treat
the model from the decay point of view, showing how the
point-interaction eigenfunctions dissipate due to the tunneling
between the points and the line; in the appendix we demonstrate
that in the weak-coupling case the decay is approximately
exponential. The main result is stated and proved in Sec.~6.

\section{Quantum Zeno dynamics}

Following general principles of quantum decay kinematics
\cite[Chap.~1]{Ex} we associate with an unstable system three
objects: the state Hilbert space $\HH$ describing all of its
states including the decayed ones, the full Hamiltonian $H$ on
$\HH$ and the projection $P$ which specifies the subspace of
states of the unstable system alone. $H$ is, of course, a
self-adjoint operator, we need to assume that it is \emph{bounded
from below.}

The question about the existence of Zeno dynamics mentioned above
can be then stated in this context generally as follows: does the
limit
 \begin{equation} \label{prod}
 (P \e^{-iHt/n} P)^n \longrightarrow \e^{-iH_Pt}
 \end{equation}
hold as $n\to\infty$, in which sense, and what is in such a case
the operator $H_P$? Let us start from the end and consider the
quadratic form $u \mapsto \|H^{1/2}Pu\|^2$ with the form domain
$D(H^{1/2}P)$ which is closed but in general it may not be densely
defined. The classical results of Chernoff \cite{Ch1, Ch2} suggest
that the operator associated with this form, $H_P :=
(H^{1/2}P)^*(H^{1/2}P)$, is a natural candidate for the generator
of the Zeno dynamics, and the counterexamples mentioned in the
introduction show that the limit may not exist if $H_P$ is not
\emph{densely defined,} so we adopt this assumption.

 \begin{rem}
Notice that the operator $H_P$ is an extension of $PHP$, but in
general a nontrivial one. This can be illustrated even in the
simplest situation when $\dim P=1$, because if $H$ is unbounded
$D(H)$ is a proper subspace of $D(H^{1/2})$. Take $\psi_0 \in D
(H^{1/2})\setminus D(H)$ such that $H^{1/2}\psi_0$ is nonzero, and
let $P$ refer to the one-dimensional subspace spanned by $\psi_0$.
This means that $PHP$ cannot be applied to any nonzero vector
$\psi\: (=\alpha\psi_0)$ of $P\HH$ while $H_P\psi$ is well defined
and nonzero.
 \end{rem}

\noindent It is conjectured that formula (\ref{prod}) will hold
under the stated assumptions in the strong operator topology.
Proof of this claim remains an open question, though. The best
result to the date \cite{EI} establishes the convergence in a
weaker topology which includes averaging of the norm difference
with respect to the time variable. While this may be sufficient
from the viewpoint of physical interpretation, mathematically the
situation is unsatisfactory, since other results available to the
date require modifications at the left-hand side of (\ref{prod}),
either by replacement of the exponential by another Kato function,
or by adding a spectral projection  interpreted as an additional
energy measurement -- see \cite{EINZ} for more details.

There is one case, however, when the formula can be proven, namely
the situation when $\dim P<\infty$ and the density assumption
simply means that $P\HH\subset \QQ(H)$, where $\QQ(H)$ is the form
domain of $H$. Since this exactly what we
need for our example, let us state the result.

 \begin{thm} \label{zeno-fin}
Let $H$ be a self-adjoint operator in a separable Hilbert space
$\HH$, bounded from below, and $P$ a finite-dimensional orthogonal
projection on $\HH$. If $P\HH\subset \QQ(H)$, then for any
$\psi\in\HH$ and $t\ge 0$ we have
 \begin{equation} \label{prod-fin}
 \lim_{n\to\infty} (P \e^{-iHt/n} P)^n \psi = \e^{-iH_Pt}\psi\,,
 \end{equation}
uniformly on any compact interval of the time variable $t$.
 \end{thm}

 \begin{proof}
The claim can be proved in different ways, see \cite{EI} and
\cite{EINZ}. Here we use another argument the idea of which was
suggested  by G.M.~Graf and A.~Guekos \cite{GG}. Notice first that
without loss of generality we may suppose that $H$ is strictly
positive, i.e. $H \geq \delta I$ for some positive number
$\delta$. The said argument is then based on the observation that
 \begin{equation} \label{aux1}
 \lim_{t\to 0}\: t^{-1}\left\| P e^{-iHt}P
 - P\e^{-itH_Pt}P \right\| = 0
 \end{equation}
implies $\left\| (P e^{-iHt/n}P)^n - P e^{-iH_Pt} \right\| = n\,
o(t/n)$ as $n\to\infty$ by means of a natural telescopic estimate.
To establish (\ref{aux1}) one has first to check that
 $$ 
 t^{-1} \left[ (\phi, P e^{-iHt}P\psi) - (\phi,\psi)
 -it (H^{1/2}P\phi, H^{1/2}P\psi) \right] \to 0
 $$ 
as $t\to 0$ for all $\phi,\psi$ from $D(H^{1/2}P)$ which coincides
in this case with $P\HH\oplus (I-P)\HH$ by the closed-graph
theorem. The last expression is equal to
 $$ 
 \left( H^{1/2}P\phi, \left[ \frac{e^{-iHt}-I}{Ht} -i \right]
 H^{1/2}P\psi \right)
 $$ 
and the square bracket tends to zero strongly by the functional
calculus, which yields the sought conclusion. In the same way we
find that
 $$ 
 t^{-1} \left[ (\phi, P e^{-iH_Pt}P\psi) -
(\phi,\psi) -it (H_P^{1/2}\phi, H_P^{1/2}\psi) \right] \to 0
 $$ 
holds as $t\to 0$ for any vectors $\phi,\psi\in P\HH$. Next we
note that $(H_P^{1/2}\phi, H_P^{1/2}\psi) = (H^{1/2}P\phi,
H^{1/2}P\psi)$ by definition, and consequently, the expression
contained in (\ref{aux1}) tends to zero weakly as $t\to 0$,
however, in a finite dimensional $P\HH$ the weak and operator-norm
topologies are equivalent.
\end{proof}

 \begin{rem}
It is clear that the finite dimension of $P$ is essential for the
proof. The same results holds for the backward time evolution,
$t\le 0$. Moreover, the formula (\ref{prod-fin}) has non-symmetric
versions with the operator product replaced with $(P
\e^{-iHt/n})^n$ and $(\e^{-iHt/n} P)^n$ tending to the same limit
-- see \cite{EI}.
 \end{rem}

\section{A model of leaky line and dots} \label{model}

Before coming to the proper decay problem let us describe the
general setting of the model. We will consider a generalized
Schr\"odinger operator in $L^2\equiv L^2(\R^2)$ with a singular interaction
supported by a set consisting of two parts. One is a straight
line, the other is a finite family of points situated in general
outside the line, hence formally we can write our Hamiltonian as
 \begin{equation} \label{forHam}
 -\Delta -\alpha \delta (x-\Sigma ) +\sum_{i=1}^n \tilde\beta_i
 \delta(x-y^{(i)})\,,
 \end{equation}
where $\alpha>0$, $\,\Sigma :=\{(x_1,0);\,x_1\in\R^2\}$, and
$\Pi:= \{y^{(i)}\}_{i=1}^n \subset \R^2 \setminus \Sigma$. The
formal coupling constants of the two-dimensional $\delta$
potentials are marked by tildes because they are not identical
with the proper coupling parameters $\beta_i$ which define these
point interaction by means of appropriate boundary conditions.

Following the standard prescription \cite{AGHH2} one can define
the operator rigorously \cite{EK} by introducing appropriated
boundary conditions on $\Sigma \cup \Pi$. Consider functions
$\psi\in W^{2,2}_{\mathrm{loc}}(\R^{2}\setminus (\Sigma \cup \Pi
)) \cap L^{2}$ which are continuous on $\Sigma$. For a small
enough $\rho>0$ the restriction $\psi\upharpoonright
_{\mathcal{C}_{\rho ,i}}$ to the circle $\mathcal{C}_{\rho
,i}\equiv \mathcal{C}_{\rho}(y_{i}):=\{q\in \R^2:|q-y^{(i)}|=\rho
\}$ is well defined; we will say that $\psi$ belongs to
$D(\dot{H}_{\alpha , \beta })$ \emph{iff}
$(\partial^2_{x_1}+\partial^2_{x_2})\psi$ on $\R^{2}\setminus
(\Sigma \cup \Pi )$ in the sense of distributions belongs to $L^2$
and the limits
 \begin{eqnarray*} \label{}
 && \Xi_{i}(\psi):=-\lim _{\rho \rightarrow 0}\frac {1}{\ln \rho
 }\,\psi \upharpoonright _{\mathcal{C}_{\rho ,i}}\,,\;
 \Omega_{i}(\psi):=\lim _{\rho \rightarrow 0}[\psi\upharpoonright
 _{\mathcal{C}_{\rho ,i}} +\Xi_{i}(\psi)\ln \rho ] \,,\;\:
 i=1,\dots,n\,, \\ &&
 \Xi_{\Sigma }(\psi)(x_{1}):=
 \partial_{x_{2}} \psi(x_{1},0+)-
 \partial_{x_{2}}\psi(x_{1},0-)\,, \quad
 \Omega_{\Sigma }(\psi)(x_{1}):= \psi(x_{1},0)
 \end{eqnarray*}
exist, they are finite, and satisfy the relations
\begin{equation}  \label{boucon}
2\pi \beta _i \Xi _{i}(\psi)=\Omega  _{i}(\psi)\,, \quad \Xi
_{\Sigma}(\psi)(x_{1})=-\alpha \Omega  _{\Sigma }(\psi)(x_{1})\,,
\end{equation}
where $\beta _{i}\in \R$ are the true coupling parameters; we put
$\beta \equiv (\beta _1 ,\dots,\beta _n )$ in the following. On
this domain we define the operator $\dot H_{\alpha, \beta}:D(\dot
H_{\alpha, \beta})\rightarrow L^{2}$ by
$$ \dot H_{\alpha, \beta}\psi(x)=-\Delta \psi(x)\quad \mathrm{for}\quad
x\in \R^{2} \setminus (\Sigma \cup \Pi)\,. $$
It is now a standard thing to check that $\dot H_{\alpha, \beta}$
is essentially self-adjoint \cite{EK}; we identify its closure
denoted as $H_{\alpha, \beta}$ with the formal Hamiltonian
(\ref{forHam}).

To find the resolvent of $H_{\alpha ,\beta }$ we start from
$R(z)=(-\Delta -z)^{-1}$ which is for any $z\in \C\setminus
[0,\infty)$ an integral operator with the kernel
$G_{z}(x,x')=\frac{1}{2\pi}K_{0} (\sqrt{-z}|x-x'|)$, where $K_{0}$
is the Macdonald function and $z\mapsto \sqrt{z}$ has
conventionally a cut at the positive halfline; we denote by
$\mathbf{R}(z)$ the unitary operator with the same kernel acting
from $L^{2}$ to $W^{2,2}\equiv W^{2,2}(\R^2)$. We introduce
two auxiliary Hilbert
spaces, $\mathcal{H}_0:= L^{2}(\mathbb{R})$ and $\mathcal{H}_1:=
\mathbb{C}^n$, and the corresponding trace maps $\tau
_j:W^{2,2}\to \mathcal{H}_j$ which act as
 $$
 \tau _0  \psi:=\psi\!\upharpoonright_{\,\Sigma }\,, \quad \tau_1
 \psi:=\psi\!\upharpoonright_{\,\Pi}=(\psi\!\upharpoonright _{\,\{y^{(1)}\}},
 \dots,\psi\!\upharpoonright _{\,\{y^{(n)}\}})\,,
 $$
respectively; they allow us to define the canonical embeddings of
$\mathbf{R} (z)$ to $\mathcal{H}_{i}$, i.e.
 $$ 
 \mathbf{R}_{iL}(z)=\tau _{i}R(z):L^{2}\rightarrow \mathcal{H}
 _{i}\,,\quad \mathbf{R}_{Li}(z)=[\mathbf{R}_{iL}(z)]^{\ast
 }:\mathcal{H} _{i}\rightarrow L^{2}\,,
 $$ 
and $\mathbf{R}_{ji}(z)=\tau _{j}\mathbf{R}_{Li}(z):\mathcal{H}%
_{i}\rightarrow \mathcal{H}_{j}$, all expressed naturally through
the free Green's function in their kernels, with the variable
range corresponding to a given $\mathcal{H}_{i}$. The
operator-valued matrix $\Gamma (z)=[\Gamma
_{ij}(z)]:\mathcal{H}_{0}\oplus\mathcal{H}_{1}\rightarrow
\mathcal{H}_{0}\oplus\mathcal{H}_{1}$ is defined by
 \begin{eqnarray*} \label{forg22}
 \Gamma _{ij}(z)g &\!:=\!& -\mathbf{R}_{ij}(z)g \qquad
 \mathrm{for}\quad i\neq j \quad \mathrm{and }\;\; g\in
 \mathcal{H}_{j}\,, \\ \Gamma_{00}(z)f &\!:=\!& \left[\alpha^{-1}
 -\mathbf{R}_{00}(z)\right] f \qquad \mathrm{if} \;\;
 f\in \mathcal{H}_0\,, \\ \Gamma _{11}(z)\varphi &\!:=\!&
 \left[ s_{\beta _{l}}(z) \delta_{kl} - G_{z}(y^{(k)},y^{(l)}) (1\!-\!
 \delta_{kl}) \right]_{k,l=1}^{n} \varphi \quad \mathrm{for}
 \;\; \varphi \in \mathcal{H}_1\,,
 \end{eqnarray*}
where $s_{\beta_{l}}(z)= \beta_{l}+s(z):=\beta
_{l}+\frac{1}{2\pi}(\ln \frac{\sqrt{z}} {2i}-\psi(1))$ and
$-\psi(1)$ is the Euler number.

For $z$ from $\rho(H_{\alpha ,\beta })$ the operator $\Gamma (z)$
is boundedly invertible. In particular, $\Gamma _{00}(z)$ is
invertible and it makes sense to define $D(z)\equiv
D_{11}(z):\mathcal{H}_{1}\rightarrow \mathcal{H}_{1}$ by
 \begin{equation} \label{Gamhat}
 D(z)=\Gamma _{11}(z)-\Gamma _{10}(z)\Gamma _{00}(z)^{-1}\Gamma
 _{01}(z)
 \end{equation}
which we call the \emph{reduced determinant} of $\Gamma $; it
allows us to write the inverse of $\Gamma (z)$ as
$[\Gamma(z)]^{-1}: \mathcal{H}_{0}\oplus\mathcal{H}_{1}\rightarrow
\mathcal{H}_{0}\oplus\mathcal{H}_{1}$ with the ``block elements''
defined by
 \begin{eqnarray*}
 \left[\Gamma(z)\right]_{11}^{-1} &=& D(z)^{-1}\,, \\
 \left[\Gamma(z)\right]_{00}^{-1} &=&
 \Gamma_{00}(z)^{-1} + \Gamma_{00}(z)^{-1} \Gamma_{01}(z)D(z)^{-1}
 \Gamma_{10}(z)\Gamma_{00}(z)^{-1}\,, \\
 \left[\Gamma(z)\right]_{01}^{-1} &=& -\Gamma_{00}(z)^{-1}
 \Gamma _{01}(z) D(z)^{-1}\,, \\
 \left[\Gamma(z)\right]_{10}^{-1} &=& -D(z)^{-1}
 \Gamma_{10}(z)\Gamma_{00}(z)^{-1}\,;
\end{eqnarray*}
in the above formulae we use notation $\Gamma _{ij}(z)^{-1}$ for
the inverse of $\Gamma _{ij}(z)$ and $[\Gamma (z)]_{ij}^{-1}$ for
the matrix element of $[\Gamma (z)]^{-1}$.

Before using this to express $R_{\alpha ,\beta }(z)\equiv
(H_{\alpha ,\beta }-z)^{-1}$ we introduce another notation which
allow us to write $R_{\alpha ,\beta }(z)$ through a perturbation
of the ``line only'' Hamiltonian $\tilde{H}_{\alpha }$ the
resolvent of which is the integral operator
 $$ 
 R_{\alpha }(z)=R(z)+R_{L0}(z)\Gamma ^{-1}_{00}R_{0L}(z)
 $$ 
for $z\in \C\setminus [-\frac{1}{4}\alpha ^{2},\infty )$. We
define $\mathbf{R}_{\alpha ; L1}(z):\mathcal{H}_{1}\rightarrow
L^{2}$ and $\mathbf{R}_{\alpha ;1L}(z):L^2 \rightarrow
\mathcal{H}_{1}$ by
 $$ 
 \mathbf{R}_{\alpha ; 1L}(z)\psi:=R_{\alpha }(z)
 \psi\upharpoonright _{\Pi}
 \quad \mathrm{for}\;\;  \psi\in L^2
 $$ 
and $\mathbf{R}_{\alpha ;L1}(z):=\mathbf{R}^{\ast }_{\alpha ;
1L}(z)$; the resolvent difference between $H_{\alpha ,\beta }$ and
$\tilde{H}_{\alpha }$ is given then by Krein's formula. Now we can
state the result; for the proof and a more detailed discussion we
refer to \cite{EK}.

 \begin{thm} \label{resoth}
 For any $z\in \rho (H_{\alpha ,\beta })$ with $\mathrm{Im \,}z>0$
 we have
 $$ 
 R_{\alpha ,\beta }(z) =R(z)+\sum_{i,j=0}^{1} \mathbf{R}
 _{Li}(z)[\Gamma(z)]_{ij}^{-1}\mathbf{R}_{jL}(z) =R_{\alpha}(z)
 + \mathbf{R}_{\alpha;L1}(z) D(z)^{-1} \mathbf{R}_{\alpha;1L}(z)\,.
 $$ 
\end{thm}

\noindent These formul{\ae} make it possible to analyze spectral
properties of the operator $H_{\alpha ,\beta }$, see again
\cite{EK} for more details. In this paper we will be concerned
with one aspect of this problem only, namely with perturbations of
embedded eigenvalues.

\section{Resonance poles} \label{s: respole}

The decay in our model is due to the tunneling between the points
and the line. This interaction is ``switched off'' if the line is
removed (formally speaking, put to an infinite distance).
Consequently, the free Hamiltonian from the decay point of view is
the point interaction only $\tilde{H}_{\beta}:= H_{0,\beta}$.
Depending on the configuration of the set $\Pi$ and the coupling
parameters $\beta$ this operator has $m$ eigenvalues, $1\le m\le
n$. \emph{We will always assume} in the following that they
satisfy the condition
 \begin{equation} \label{hypoth1}
 -\frac14 \alpha^2< \epsilon_1< \cdots< \epsilon_m <0 \quad
 \mathrm{and} \quad m>1\,,
 \end{equation}
i.e., the discrete spectrum of $\tilde{H}_{\beta}$ is simple,
contained in (the negative part of) $\sigma(\tilde{H}_{\alpha})=
\sigma _{ac}(\Hab)=(-\alpha ^2 /4 ,\infty )$, and consists of more
than a single point. Let us specify the interactions sites by
their Cartesian coordinates, $y^{(i)}=(c_i ,a_i )$. We also
introduce the notations $a=(a_1 ,...,a_n )$ and
$d_{ij}=|y^{(i)}-y^{(j)}|$ for the distances between point
interactions.

To find resonances in our model we will rely on a Birman-Schwinger
type argument \footnote{We will follow here the idea which was
precisely discussed in \cite{EK}}. More specifically, our aim is
to find poles of the resolvent through zeros of the
operator-valued function (\ref{Gamhat}). First we have to find a
more explicit form of $D(\cdot )$; having in mind that resonance
poles have to be looked for on the second sheet we will derive the
analytical continuation of $D(\cdot )$ to a subset $\Omega_-$ of
the lower halfplane across the segment $(-\alpha ^2 /4 , 0 )$ of
the real axis; for the sake of definiteness we employ the notation
$D(\cdot)^{(l)}$ where $l=-1,0,1$ refers to the argument $z$ from
$\Omega_-$, the segment $(-\alpha ^2 /4 , 0 )$, and the upper
halfplane, $\mathrm{Im\,}z>0$, respectively. Using the resolvent
formula of the previous section we see that the first component of
the operator $\Gamma_{11}(\cdot)^{(l)}$ is the $n\times n $ matrix
with the elements
 $$
 \Gamma _{11;jk}(\cdot)^{(l)}=-\frac{1}{2\pi}
 K_{0}(d_{jk}\sqrt{-\cdot})\quad \mathrm{for}\; j\neq k
 $$
and
 $$\Gamma_{11;jj}(\cdot)^{(l)}= \beta_{j}+1/2\pi
 (\ln \sqrt{(-\cdot)}-\psi (1))
 $$
for every $l$. To find an explicit form of the second component
let us introduce
 $$
 \mu_{ij}(z,t):=\frac{i\alpha }{2^{5}\pi }\frac{(\alpha
 -2i(z-t)^{1/2})\, \e^{i(z-t)^{1/2}(|a_{i}|+|a_{j}|)}}{t^{1/2}(z-t)^{1/2}}
 \,\e^{it^{1/2}(c_i -c_j )}
 $$
and $\mu_{ij}^{0}(\lambda ,t):= \lim_{\eta\to0+}\mu_{ij}(\lambda
+i\eta,t)$ cf.~\cite{EK}. Using this notation we can rewrite the matrix elements
of $(\Gamma _{10}\Gamma _{00}^{-1}\Gamma _{01})^{(\cdot)}(\cdot)$
in the following form,
 \begin{eqnarray*}
 \theta_{ij}^{(0)}(\lambda ) &\!=\!& \mathcal{P}\int_{0}^{\infty}
 \frac{\mu_{ij}^0 (\lambda , t) }{t-\lambda -\alpha^{2}/4}\,\mathrm{d}t
 +g_{\alpha,ij}(\lambda )\,, \qquad\;\: \lambda \in (-\frac{\alpha ^2}{4},0)
 \\
 \theta_{ij}^{(l)}(z) &\!=\!& l \int_{0}^{\infty}\frac{\mu _{ij}(z,t)}
 {t-z -\alpha^{2}/4}\,\mathrm{d}t+(l-1) g_{\alpha,ij}(z )\quad
 \mathrm{for} \;\; l=1,\, -1
 \end{eqnarray*}
where $\mathcal{P}$ means the principal value and
 $$
 g_{\alpha,ij}(z):= \frac{i\alpha }{(z+\alpha ^2/4)^{1/2}}\,
 \e^{-\alpha(|a_{i}|+|a_j |)/2}\, \e ^{i(z+\alpha^{2}/4)^{1/2}
 (c_i -c_j )}\,.
 $$
Proceeding in analogy with \cite{EK} we evaluate the determinant
of $D(\cdot)^{(\cdot)}$ as
 $$
 d(z)^{(l)}\equiv d(a,z)^{(l)}=\sum_{\pi \in \mathcal{P}_{n}}
 \mathrm{sgn\,}\pi \left( \sum _{j=1}^{n}(-1)^{j}
 (S^{j}_{p_1 ,\dots,p_n })^{(l)}
 +\Gamma _{11;1p_1}\dots\Gamma_{11;np_n}\right)(z)\,,
 $$
where $\mathcal{P}_{n}$ denotes the permutation group of $n$
elements, $\pi=(p_1,\dots,p_n)$, and
 $$
 (S^{j}_{p_1 ,...,p_n })^{(l)}
 =\theta _{jp_{1}}^{(l)}A^{j}_{p_2,\dots,p_n }
 $$
with
 $$
 A^{j}_{i_2 ,...,i_n }:=\left\{
 \begin{array}{lcl}
 \Gamma_{11;1i_2}\dots\Gamma_{11;j-1,i_j}\Gamma_{11;j+1,i_{j+1}}
 \dots\Gamma_{11;ki_k} &\quad \mathrm{if}\;\; & j>1
 \\ \Gamma_{11;2i_2}\dots\Gamma_{11;ki_k}&\quad \mathrm{if}\;\; & j=1
 \end{array}\right .
 $$
After this preliminary we want to find roots of the equation
$d(a,z)^{(l)}(z)=0$. On a heuristic level the resonances are due
to tunneling between the line and the points, thus it is
convenient to introduce the following reparametrization,
 $$
 \tilde{b}(a)\equiv (b_1(a),\dots, b_n (a))\quad b_{i}(a)
 =\e^{-|a_i|\sqrt{-\epsilon_i}}\,
 $$
and to put $\eta(\tilde{b},z)=d^{(-1)}(a,z)$. As we have said the
absence of the straight-line interaction can be regarded in a
sense as putting the line to an infinite distance from the points,
thus corresponding to $\tilde{b}=0$. In this case we have
$$\eta (0,z)=\sum_{\pi \in \mathcal{P}_{n}}\mathrm{sgn\,}\pi \left(
\Gamma _{11;1p_1}\dots\Gamma _{11;np_n}\right)(z) = \det
\Gamma_{11}(z)\,,$$
so the roots of the equation $\eta (0,z)=0$ are nothing else than
the eigenvalues of the point-interaction Hamiltonian $\tilde
H_\beta$; with the condition (\ref{hypoth1}) in mind we have
 $$
 \eta (0,\epsilon_i )=0\,, \quad i=1,...,m\,.
 $$
Now one proceeds as in \cite{EK} checking that the hypotheses of
the implicit-function theorem are satisfied; then the equation
$\eta(\tilde{b},z)=0$ has for all the $b_i$ small enough just $m$
zeros which admit the following weak-coupling asymptotic
expansion,
\begin{equation}\label{poles}
z_i(b)=\epsilon_i +\mathcal{O}(b)+i\mathcal{O}(b)\quad
\mathrm{where}\quad  b:=\max _{1\leq i\leq m}b_i \,.
\end{equation}

 \begin{rem} \label{symm}
 If $n\ge 2$ there can be eigenvalues of $\tilde H_\beta$ which
 remain embedded under the line perturbation due to a symmetry;
 the simplest example is a pair of point interactions with the
 same coupling and mirror symmetry with respect to $\Sigma$. From the
 viewpoint of decay which is important in this paper they represent
 a trivial case which we exclude in the following. Neither shall we
 consider resonances which result from a slight violation of such a
 symmetry -- cf. a discussion in \cite{EK}.
 \end{rem}

\section{Decay of the dot states}

As usual the resonance poles discussed above can be manifested in
two ways, either in scattering properties, here of a particle
moving along the ``wire'' $\Sigma$, or through the time evolution
of states associated with the ``dots'' $\Pi$. By assumption
(\ref{hypoth1}) there is a nontrivial discrete spectrum of $\tilde
H_\beta$ embedded in $(-\frac14 \alpha^2,0)$. Let us denote the
corresponding normalized eigenfunctions $\psi_j\,,\: j=1,\dots,m$,
given by
 \begin{equation} \label{efs}
 \psi_j(x) =\sum_{i=1}^m d_i^{(j)} \phi_i^{(j)}(x)\,, \quad
 \phi_i^{(j)}(x):= \sqrt{-\frac{\epsilon_j}{\pi}}\,
 K_{0} (\sqrt{-\epsilon_j}|x-y^{(i)}|)
 \end{equation}
in accordance with \cite[Sec.~II.3]{AGHH2}, where the vectors
$d^{(j)}\in\C^m$ satisfy the equation
 \begin{equation} \label{dvect}
 \Gamma_{11}(\epsilon_j)d^{(j)}=0
 \end{equation}
and a normalization condition which in view of $\|\phi_i^{(j)}\|
=1$ reads
 \begin{equation} \label{dnorm}
 |d^{(j)}|^2 +2\mathrm{Re\,} \sum_{i=2}^m \sum_{k=1}^{i-1}\,
 \overline{d_i^{(j)}} d_k^{(j)} (\phi_i^{(j)},\phi_k^{(j)})=1\,.
 \end{equation}
In particular, if the distances between the points of $\Pi$ are
large (the natural length scale is given by
$(-\epsilon_j)^{-1/2}$), the cross terms are small and $|d^{(j)}|$
is close to one.

Let us now \emph{specify the unstable system} of our model by
identifying its state Hilbert space $P\HH$ with the span of the
vectors $\psi_1,\dots,\psi_m$. Suppose that it is prepared at the
initial instant $t=0$ at a state $\psi\in P\HH$, then the decay
law describing the probability of finding the system undecayed at
a subsequent measurement performed at $t$, without disturbing it
in between \cite{Ex}, is
 \begin{equation} \label{decay}
 P_\psi(t) = \| P \e^{-iH_{\alpha, \beta}t}\psi\|^2.
 \end{equation}
We are particularly interested in the \emph{weak-coupling
situation} where the distance between $\Sigma$ and $\Pi$ is a
large at the scale given by $(-\epsilon_m)^{-1/2}$. Since our
model bears resemblance with the (multidimensional) Friedrichs
model one can conjecture in analogy with \cite{De} that the
leading term in $P_\psi(t)$ will come from the appropriate
semigroup evolution on $P\HH$, in particular, for the basis states
$\psi_j$ we will have a dominantly exponential decay,
$P_{\psi_j}(t) \approx \e^{-\Gamma_j t}$ with $\Gamma_j =
2\,\mathrm{Im\,}z_j(b)$. A precise discussion of this question is
postponed to appendix -- see Sec.~7 below.

 \begin{rem} \label{lifetime}
 The quantities $\Gamma_j^{-1}$ provide thus a natural time scale
 for the decay and we will use $\max_j \Gamma_j^{-1}$ as a measure
 of the system lifetime. A \emph{caveat} is needed, however, with
 respect to the notion of lifetime \cite{Ex} which is conventionally
 defined as $T_\psi = \int_0^\infty P_\psi(t)\,\D t$. It has been
 shown in \cite{EK} that $P\HH$ is not contained is the absolutely
 continuous subspace of $H_{\alpha, \beta}$ if $n=1$, and the argument
 easily extends to any $n\in\N_0$. This means that a part of the
 original state survives as $t\to\infty$, even if it is a small one in
 the weak-coupling case. It is a long-time effect, of course, which
 has no relevance for the problem considered here.
 \end{rem}

\section{Stable and Zeno dynamics in the model}

Suppose now finally that we perform the Zeno time at our decaying
system characterized by the operator $H_{\alpha, \beta}$ and the
projection $P$. The latter has by assumption the dimension $1<m
<\infty$ and it is straightforward to check that $P\HH \subset \QQ
(\Hab)$. Moreover the form associated with generator $H_P$ has in
the quantum-dot state basis the following matrix representation
 \begin{equation} \label{matrix}
 (\psi_j, H_P\psi_k) = \delta_{jk} \epsilon_j
 -\alpha \int_\Sigma \bar\psi_j(x_1,0) \psi_k(x_1,0)\, \D x_1\,,
 \end{equation}
where the first term corresponds, of course, to the ``dots-only''
operator $\tilde H_\beta$.

 \begin{thm} \label{compar}
 The two dynamics do not differ significantly for times satisfying
 \begin{equation} \label{timeest}
 t\ll C\,\e ^{2\sqrt{-\epsilon}|\tilde{a}|}\,,
 \end{equation}
 where $C$ is a positive constant and $|\tilde{a}|=\min_{i} |a_i
 |$, $\epsilon = \max_{i}\epsilon_i $.
 \end{thm}

 \begin{proof}
The difference is characterized by the operator $\UU_t:= (\e^{-i
\tilde H_\beta t}- \e^{-iH_Pt})P$. Taking into account the
unitarity of its parts together with a functional calculus
estimate based on $|\e^{iz}\!-1|\le |z|$ we find that the norm of
$\UU_t$ remains small as long as $t\|(\tilde H_\beta - H_P)P\| \ll
1$. Thus to check (\ref{timeest}) we have to estimate norm of the
operator $(\tilde H_\beta - H_P)P$ acting in $P\mathcal{H}$; in
the basis of the vectors $\{\psi _{j}\}_{j=1}^{m}$ it is
represented by $m\times m$ matrix with the elements
 $$ s_{ij}= \alpha (\psi_i ,\psi_j)_{\Sigma}\,, $$
where $(\psi_i,\psi_j)_{\Sigma} := \int_\Sigma \bar\psi_i(x_1,0)
\psi_j(x_1,0)\, \D x_1$. Using the representation (\ref{efs}) we
obtain
 $$
 s_{ij}= \alpha \sum_{(l,k)\in M\times M}
 \bar{d}_{l}^{(i)}d_{k}^{(j)}(\phi^{(i)}_{l}
 ,\phi^{(j)}_{k})_{\Sigma}
 $$
where $M$ is a shorthand for $(1,...,m)$. To proceed further we
use Schur-Holmgren bound by which the norm of $(\tilde H_\beta -
H_P)P$ does not exceed $mS$, where $S:= \max_{(i,j)\in M\times
M}|s_{ij}|$, and the last named quantity can be estimated by
 $$
 S\leq \alpha m^2 \max_{(i,j,k,l)\in M^4}
 |\bar{d}_{l}^{(i)}d_{j}^{(k)}(\phi^{(i)}_{l},
 \phi^{(j)}_{k})_{\Sigma}|\,.
$$
The final step is to estimate the expressions $(\phi^{(i)}_{l}
,\phi^{(j)}_{k})_{\Sigma}$. Using the momentum representation of
Macdonald function we obtain
 $$
 (\phi^{(i)}_{l} ,\phi^{(j)}_{k})_{\Sigma}=\frac{\sqrt{\epsilon_{i}
 \epsilon_{j}}}{2}\int_{\R}\frac{\e^{-((p_{1}^{2}
 -\epsilon_{i})^{1/2}|a_{l}|-(p_{1}^{2}-
 \epsilon_{j})^{1/2}|a_{k}|)}}{(p_{1}^{2}-\epsilon_{i})^{1/2}
 (p_{1}^{2}-\epsilon_{j})^{1/2}}\,\e^{ip_{1}(c_{k}-c_{l})}\,
 \mathrm{d}p_{1}\,,
$$
where $y^{(i)}=(c_{i}, a_{i})$ as before. A simple estimate of the
above integral yields
 $$
 (\phi^{(i)}_{l} ,\phi^{(j)}_{k})_{\Sigma}\leq \frac{\pi }{2}\,
 \frac{\epsilon_{min}}{\sqrt{-\epsilon}}\, \e^{-2\sqrt{-\epsilon}|a|}
 $$
where $\epsilon_{min}=\min _{i} \epsilon_{i},\:
|\tilde{a}|=\min_{i} |a_i |$, and $\epsilon =\max_{i} \epsilon_i$.
In conclusion, we get the bound
 $$
 \|(\tilde H_\beta - H_P)P\| \leq C\e ^{-2\sqrt{-\epsilon}|a|}\,,
 $$
where $C:=\frac12 \pi m^3 \alpha\, \epsilon_{min}
(-\epsilon)^{-1/2} \max_{(i,j,k,l)\in M^4 } |\bar{d}_{l}^{(i)}
d_{j}^{(k)}|$.
 \end{proof}

\section{Appendix: pole approximation for the decaying states}

Let us now return to the claim that the decay is approximately
exponential when the distances of the points from the line are
large. Let $\psi_j$ be the $j$-th eigenfunction of the
point-interaction Hamiltonian $\tilde H_\beta$ with the eigenvalue
$\epsilon_j$; the related one-dimensional projection will be
denoted $P_j$. Then we make the following claim.
\begin{thm}\label{tdecay1}
Suppose that $\Hab$ has no embedded eigenvalues. Then in the limit
$b\to 0$ where $b$ is defined in (\ref{poles}) we have, pointwise
in $t\in(0,\infty)$,
 $$
 \|P_j \e^{-i\Hab t}\psi_j -\e^{-iz_j t}\psi_j \|\to 0\,.
 $$
\end{thm}
\bigskip

\noindent To prove the theorem we need some preliminaries. For
simplicity, we denote $U_t (\epsilon):= \e^{-i\epsilon t}$ for a
fixed $t>0$. It was shown in \cite{EK} that the operator $\Hab$
has at least one and at most $n$ isolated eigenvalues. We denote
them by $\epsilon _{\alpha \beta , k},\: k=1,...,l$ with $l\leq
n$, and use $\psi _{\alpha \beta , k}$ as symbols for the
corresponding (normalized) eigenfunctions. Then the spectral
theorem gives
 \begin{equation}\label{evolution}
 P_j \,\e^{-i\Hab t}\psi_j=\sum_{k=1}^{m}U_{t}(\eab)|(\psi_j,
 \psi_{\alpha \beta , k})|^2\psi_j + P_j \int_{-\alpha^2/4}^\infty
 U_{t}(\lambda )\mathrm{d}E(\lambda )\psi_j\,,
\end{equation}
where $E(\cdot )\equiv E_{\alpha , \beta }(\cdot )$ is the
spectral measure of $\Hab$. By assumption there are no embedded
eigenvalues (cf.~Remark~\ref{symm}) and by \cite{EK} also the
singularly continuous component is void, hence the second term is
associated solely with $\sigma_\mathrm{ac}(\Hab)$. Let us first
look at this contribution to the reduced evolution. The key
observation is that one has a spectral concentration in the set
$\triangle_ \varepsilon \equiv \triangle_ \varepsilon (b):=
(\epsilon_j- \varepsilon (b), \epsilon_j+ \varepsilon (b))$ with a
properly chosen $\varepsilon (b)$; we denote its complement as
$\bar{\triangle}_ \varepsilon := \sigma _\mathrm{ac} (\Hab)
\setminus \triangle_ \varepsilon$.

\begin{lem} \label{ldecay1}
Suppose that $\varepsilon(b)\to 0$ and $\varepsilon(b)^{-1}b\to 0$
holds as $b\to 0$, then we have
$$
\|P_j \int_ {\bar{\triangle}_ \varepsilon}U_{t}(\lambda
)\mathrm{d}E(\lambda )\psi_j\|\to 0\,.
$$
\end{lem}
\begin{proof}
Given an arbitrary Borel set $\triangle \subset \sigma
_\mathrm{ac} (\Hab )$ and a projection $P$ we have the following
simple inequality,
 \begin{equation}\label{concent1}
 \|P \int_ {\triangle}U_{t}(\lambda )\mathrm{d}E(\lambda )f\|\leq
 \|E(\triangle )f\|\,,
 \end{equation}
and another straightforward application of the spectral theorem
gives
 \begin{equation}\label{concent2}
 \|(\Hab -\epsilon _j )f\|^2 \geq \int_ {\bar{\triangle}_
 \varepsilon}|\lambda -\epsilon _j |^2 (\mathrm{d}E(\lambda
 )f,f)\geq \varepsilon (b)^2\|E(\bar{\triangle}_
 \varepsilon)f\|^2
 \end{equation}
for any $f\in D(\Hab)$. To make use of the last inequality we need
a suitable function from the domain of $\Hab$. It is clear that
one cannot use $\psi_j$ directly because it does not satisfy the
appropriate boundary conditions at the line $\Sigma$, thus we take
instead its modification $f_b =\psi_j + \phi _b $, where $\phi_b
\in L^2 (\R^2 )$ vanishes on $\Pi \cup \Sigma$ and satisfies the
following assumptions:

(a1) $\Xi _\Sigma (\phi_b )=-\alpha \Omega _\Sigma (\psi _j )$

(a2) $\|\phi_b \|=\mathcal{O}(b)$ and $\|\Delta \phi_b\| =\mathcal{O}(b)$.

\noindent In view of (\ref{boucon}) the first condition guarantees
that $f_b \in D(\Hab)$, while the second one expresses
``smallness'' of the modification. It is not difficult to
construct such a family. For instance, one can take for $\phi _b$
a family of $C^2$ functions with supports in a strip neighbourhood
of $\Sigma $ of width $d_\Sigma$ assuming that $\phi _b $ behaves
in the vicinity of $\Sigma $ as $\frac12 {\alpha} \Omega _\Sigma
(\psi _j )(x_1 )|x_2 |$. Since $|\Omega _\Sigma (\psi_j )|\leq C
b$, where $C$ is positive constant we can choose $d_\Sigma
=\mathcal{O} (b)$. Using (a1) and $(\tilde H_\beta -\epsilon_j
)\psi_j=0$ we get
 $$
 (\Hab -\epsilon_j )f_b =-\Delta \phi_b -\epsilon_j \phi_b\,,
 $$
so the condition (a2) gives
$$
 \|(\Hab -\epsilon _j)f_b\|=\mathcal{O}(b)\,.
$$
This relation together with (\ref{concent2}) yields
$\|E(\bar{\triangle}_\varepsilon)f_b\| =\mathcal{O}(b)\varepsilon
(b)^{-1}$. Combining it further with (\ref{concent1}) and using
the inequality
 $$
 \|E(\bar{\triangle}_\varepsilon)\psi_j \|\leq \|\phi _b
 \|+\|E(\bar{\triangle}_\varepsilon)f_b\|
 $$
and the condition (a2) we get the sought result.
\end{proof}

The next step is to show that the main contribution to the reduced
evolution of the unstable state comes from the interval $\triangle
_\varepsilon$.
\begin{lem}\label{ldecay2}
Under the assumptions of Lemma~\ref{ldecay1} we have
$$
\|P_j \int_ {\triangle _\varepsilon}U_t (\lambda )\mathrm
{d}E(\lambda )\psi_j -U_t (z_j)\psi_j \|\to 0
$$
for any fixed $t>0$ in the limit $b\to 0$.
\end{lem}
\begin{proof}
Let $R_{\alpha, \beta }^{\mathrm{II}}$ stand for the second-sheet
continuation of the resolvent of $\Hab$. Using the results of
Sec.~\ref{s: respole} we can write it for a fixed $j$ as
\begin{equation} \label{redres}
R_{\alpha, \beta }^{\mathrm{II}}(z)=\sum_{k=1} ^m \frac{B^
{(k)}_b}{z-z_k}+A_b(z)\,,
\end{equation}
where $B^{(k)}_b$ is a one-parameter family of rank-one operators
and $A_b(\cdot)$ is a family of analytic operator-valued functions
to be specified later. Mimicking now the argument of
\cite[Sec.~3.1]{Ex} which relies on Stone's formula and
Radon-Nikod\'ym theorem we find that the spectral-measure
derivative acts at the vector $\psi_j$ as
 \begin{equation}\label{specder}
 \frac{\mathrm{d}E(\lambda)}{\mathrm{d}
 \lambda}\,\psi_j=\left[\frac{1}{2\pi i} \sum_{k=1}^ m
 \left(\frac{(B^{(k)} _b)^\ast}{\lambda -
 \bar{z_k }}- \frac{B^{(k)}_b}{\lambda - z_k} \right) +
 \frac{1}{\pi}\, \mathrm{Im}\,A_b(\lambda )\right]\psi_j\,.
 \end{equation}
This makes it possible to estimate $P_j \int_ {\triangle
_\varepsilon}U_t (\lambda )\mathrm {d}E(\lambda )\psi_j$. Using
the explicit form of $R_{\alpha, \beta }^{\mathrm{II}}$ derived in
Sec.~\ref{s: respole} one can check that $A_b(\cdot)$ can be
bounded on a compact interval uniformly for $b$ small enough,
which means that the contribution to the integral from the last
term in (\ref{specder}) tends to zero as $\varepsilon (b)\to 0$.
The rest is dealt with by means of the residue theorem in the
usual way: we can extend the integration to the whole real line
and perform it by means of the integral over a closed contour
consisting of a real axis segment and a semicircle in the lower
halfplane, using the fact that the contribution from the latter
vanishes when the semicircle radius tends to infinity. It is clear
that only the $m$ poles in (\ref{specder}) contained in the lower
halfplane contribute, the $k$-th one giving $U_t (z_k
)P_jB_b^{(k)} \psi_j$; an argument similar to Lemma~\ref{ldecay1}
shows that the integral over $\R\setminus \triangle _\varepsilon$
vanishes as $b\to 0$, and likewise, the integral over semicircle
vanishes in the limit of infinite radius.

Furthermore, since $P_j$ is one-dimensional we have $P_jB_b^{(k)}
\psi_j= c^{(k)}_b \psi_j$ where $b\mapsto c^{(k)}_b$ are
continuous complex functions, well defined for $b$ small enough.
Hence the above discussion allows us to conclude that
 \begin{equation} \label{conaux}
 \|P_j \e ^{-i \Hab t }\psi_j -
 \sum_{k=1}^ m c^{(k)}_b \e ^{-iz_k t}\psi_j\|\to 0
 \quad \mathrm{as}\quad b\to 0\,.
\end{equation}
Our next task is show that for $k\neq j$ we have $c^{(k)}_b\to0$
as $b\to 0$ and $c_b^{(j)}\to 1$ at the same time. To this aim it
suffices to check that $B^{(k)}_b$ converges to $P_k$ for $b \to
0$. First we observe that the terms involved in the resolvent
$R_{\alpha ,\beta }$ derived in Theorem~\ref{resoth} satisfy the
following relations
$$
D(z)\to \Gamma_{11}(z)\,,\,\,\, \mathbf{R}_{\alpha;1L}(z)\to
\mathbf{R}_{1L}(z)\quad \mathrm{as}\quad b\to 0\,
$$
in the operator-norm sense; the limits are uniform on any compact
subset of the upper halfplane as well as for the analytical
continuation of $R_{\alpha ,\beta }$. Consequently, the second
component of the resolvent tends $\mathbf{R}_{L1}(z)
[\Gamma_{11}(z)]^{-1} \mathbf{R}_{1L}(z)$ which obviously has a
singular part equal to $\sum_ {k=1}^m (z-\epsilon_k )^{-1}P_k $;
this proves the claim.
\end{proof}

\noindent {\bf Proof of Theorem~\ref{tdecay1}.} In view of
(\ref{evolution}) together with Lemmata~\ref{ldecay1},
\ref{ldecay2} it remains to demonstrate that the contribution from
the discrete spectrum to (\ref{evolution}) vanishes as $b\to 0$,
i.e. that
 \begin{equation}\label{pointdecay}
 \left|\sum_{k=1}^{m}U_{t}(\eab)|(\psi_j ,
 \psi _{\alpha \beta ,k})|^2\right|\to 0\,.
 \end{equation}
This is a direct consequence of the following relation,
$$
 0=(\Hab \psi_{\alpha \beta,k}, f_b)-( \psi_{\alpha \beta,k},
 \Hab f_b)=(\epsilon_{\alpha \beta,k}-\epsilon_j )
 (\psi_{\alpha \beta,k}, f_b)+\mathcal{O}(b)\,,
$$
where $k=1,\dots,l$, and $f_b$ is the function introduced in the
proof of Lemma~\ref{ldecay1}. In combination with (\ref{hypoth1})
we get $|(\psi_j ,\psi _{\alpha \beta , k})|= \mathcal{O}(b)$
which in turn implies (\ref{pointdecay}).


\subsection*{Acknowledgment}
The research was partially supported by the ASCR and its Grant
Agency within the projects IRP AV0Z10480505 and A100480501 and by
the Polish Ministry of Scientific Research and Information
Technology under the (solicited) grant no PBZ-Min-008/PO3/2003.
Two of the authors (P.E. and S.K.) are grateful to the organizing
committee of OTAMP2004 for supporting their participation in the
conference, as well as for the hospitality at the University of
Kanazawa, where a substantial part of this work was done.

\end{document}